%% file: main.tex
\documentclass{article}
\usepackage[T1]{fontenc} 
\usepackage[utf8]{inputenc} 
\usepackage{ismir,amsmath,cite,url}
\usepackage{graphicx}
\usepackage{color}
\usepackage{mathrsfs}
\usepackage{mathptmx}

\usepackage{mathtools}
\DeclarePairedDelimiter\ceil{\lceil}{\rceil}

\usepackage[normalem]{ulem}

\title{DeepSRGM - Sequence classification and ranking in Indian Classical music with deep learning}

\threeauthors
  {Sathwik Tejaswi Madhusudhan} {University of Illinois, Urbana Champaign \\ {\tt stm4@illinois.edu}}
  {} {}
  {Girish Chowdhary} {University of Illinois, Urbana Champaign \\ {\tt girishc@illinois.edu}}

\begin{document}

\maketitle
\begin{abstract}
\input{abstract.tex}
\end{abstract}
\input{Section1.tex}

\input{Section2.tex}

\input{Section3.tex}

\input{Section4.tex}

\input{Section5.tex}

\bibliography{ISMIRtemplate}

%
%
%
%

\end{document}

%% file: abstract.tex
A vital aspect of Indian Classical Music (ICM) is \emph{Raga}, which serves as a melodic framework for compositions and improvisations  alike.  Raga Recognition  is  an  important music information retrieval task in ICM as it can aid numerous downstream applications ranging from music recommendations  to  organizing  huge  music  collections. In this work, we propose a deep learning based approach to Raga recognition. Our  approach employs efficient pre-possessing and learns temporal sequences in music data using Long Short Term Memory based Recurrent Neural Networks (LSTM-RNN). We train and test the network on smaller sequences sampled from the original audio while the final inference is performed on the audio as a whole. Our method achieves an accuracy of 88.1\% and 97 \% during inference on the \emph{Comp Music Carnatic dataset} and its 10 Raga subset respectively making it the state-of-the-art for the Raga recognition task.   Our approach also enables sequence ranking which aids us in retrieving melodic patterns from a given music data base that are closely related to the presented query sequence.

%% file: Section1.tex
\section{Introduction}\label{sec:introduction}

Carnatic and Hindustani music are the two main branches of Indian Classical Music (ICM), which underlies most of the music emanating from the Indian subcontinent. Owing to the contemplative and spiritual nature of these art forms and varied cultural influences, a lot of emphases is placed on melodic development. Raga, which governs various melodic aspects of ICM, serves as a framework for compositions and improvisations. Krishna et al. \cite{krishna2012carnatic} define a Raga as the ``collective melodic expression that consists of phraseology which is part of the identifiable macro-melodic movement''. A major portion of contemporary Indian Music including film, folk and other forms of music heavily draw inspiration from ICM \cite{ganti2013bollywood, gulati2016time}.


Numerous melodic attributes\label{Raga_Attributes} make Ragas distinctive in nature, such as the \emph{svara} (roughly, a musical note), the \emph{gamaka} (for example oscillatory movement about a given note, slide from one note to another etc \cite{koduri2012raga}), \emph{arohana} and \emph{avarohana} (upward and downward melodic movement) and melodic phrases/ motifs \cite{gulati2016time, krishna2012carnatic}. Emotion is another attribute that makes Ragas distinctive. 
Balkwill et al.\cite{balkwill1999cross} conducted studies on the perception of Ragas and observed that even an inexperienced listener is able to identify emotions portrayed by various Ragas. 
Given the importance of Raga in ICM, Machine Learning (ML) based automatic raga recognition can aid in organizing large audio libraries, labeling recording in the same, etc. 

We believe automatic raga recognition has tremendous potential in empowering content-based recommendation systems pertaining to ICM and contemporary Indian music.  However, Raga recognition is not a trivial problem. There are numerous examples where two or more Ragas have the same or a similar set of notes but are worlds apart in the musical effect they produce due factors like the gamaka, temporal sequencing (which has to abide by the constraints presented in the arohana and avarohana), as well as places of svara emphasis and rest. Raga identification is an acquired skill that requires significant training and practice. As such, automatic Raga classification methods have been widely studied. However, existing methods based on Pitch Class Distributions (PCD) disregard the temporal information and are highly error-prone, while other methods (reviewed in detail in Section~\ref{sec:related}) are highly dependent on preprocessing and hand made features, which limit the performance of such methods. 

In this work, we introduce a deep learning based solution to Raga recognition and Raga content-based retrieval. First, we reformulate the problem of Raga recognition as a sequence classification task performed using an LSTM RNN based architecture with attention. We then show that the same network, with minimal fine-tuning based on the triplet margin loss \cite{wang2014improving}, can be used for sequence ranking. We introduce sequence ranking as a new sub-task of automatic Raga recognition. In this configuration, the model can be used to perform Raga content-based retrieval. A user provides the model with a query sequence and the model returns sequences that are very closely related to the presented query. This can aid in downstream applications like recommendation systems and so on. 

Deep learning has proven to be an indispensable asset due to its flexibility, scalability,  learning abilities, end to end training, and most importantly, unprecedented success in modeling unstructured spatial and temporal data over the past few years. Our work draws inspiration from the successes in \cite{van2013deep, wang2014improving} introduce a convolutional neural network based recommendation system for music which overcomes and outperforms various drawbacks of traditional recommendation systems like collaborative filtering, a bag of words method, etc. It also leverages LSTM-RNN and attention based models which have proven to be extremely effective in tasks like sequence classification, sequence to sequence learning, neural machine translation, image captioning task, etc., all of which directly deal with sequences.


To summarize, the main contributions of our work are as follows :
\begin{itemize}
    \item We present a new approach to Raga recognition using deep learning using LSTM based RNN architecture to address problem of Raga recognition.
    \item With our approach we obtain 97.1 \% on the 10 Raga classification task and 88.1 \% accuracy on the 40 Raga classification task on the Comp Music Carnatic Music Dataset (CMD), hence improving the state-of-the-art on the same.
    \item We introduce sequence ranking as a new sub task of Raga recognition, which can be used in creating Raga content based recommendation systems.
\end{itemize}

Note that we refer to our Raga recognition (i.e sequence classification) model as SRGM1 and the sequence ranking model as SRGM2 throughout the paper.

\section{Related Works}
\label{sec:related}
Broadly speaking, ML methods for Raga classification can either be PCD based or sequence based. Many of the previous works have focused on PCD based methods for Raga recognition \cite{koduri2014intonation, dighe2013swara, dighe2013scale} with Chordia et al. \cite{chordia2013joint}, currently holding the best performing results for PCD based methods with an accuracy of 91\%. \cite{koduri2012raga} presents an in-depth study of various distribution based methods for Raga classification. Although an intuitive and an effective approach, the major shortcoming of PCD based methods is that it disregards temporal information and hence is error prone. To overcome this, previous works have used Hidden Markov Models based approach\cite{krishna2011identification}, convolutional neural networks based approach\cite{madhusdhan2018tonic}, arohana avarohana based approach \cite{shetty2009raga} among various other methods.

\cite{sridhar2009raga, dutta2015raga} employ Raga phrase-based approach for Raga recognition. \cite{gulati2016phrase} use an interesting phrase-based approach inspired by the way in which seasoned listeners identify a Raga. This method is currently the state-of-the-art on a 10 Raga subset of CMD. \cite{gulati2016time} introduces Time Delayed Melodic Surface(TDMS) which is a feature based on delay co-ordinates that is capable of describing both tonal and temporal characteristics of the melody given a song. With TDMS, the authors achieve state-of-art performance on the CMD.

While the PCD based methods ignore the temporal information, many works like assume that the audio is monophonic. Works like \cite{krishna2011identification, shetty2009raga} attempt to take advantage of the temporal information by extracting arohana and avarohana. Although arohana and avarohana are critical components to identifying a Raga, they do not contain as much information as the audio excerpt itself. Most previous works heavily rely on prepossessing and feature extraction, most of which are handcrafted, which can prove to be a limitation as it only gets harder to devise ways to extract complex and rich features.

%% file: Section2.tex
\begin{figure}
 \centerline{
 \includegraphics[width=\columnwidth]{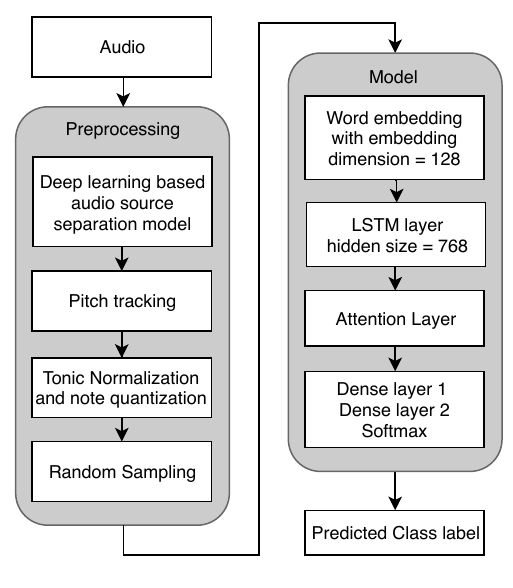}}
 \caption{Figure shows various preprocessing steps and model architecture for SRGM1 (refer Section 3)}
 \label{Figure1}
\end{figure}

\section{Raga recognition as a sequence classification problem ($\mathcal{SRGM1}$)}

Sequence classification is a predictive modeling problem where an input, a sequence over space or time, is to be classified as one among several categories. In modern NLP literature, sentiment classification is one such example, where models predict sentiments of sentences. We argue that the Raga recognition is closely related to the task of sequence classification, since any audio excerpt in ICM involving vocals/melody instruments, can be broken down as a sequence of notes spread over time. To reformulate Raga recognition as a sequence classification task, we treat the notes obtained via predominant melody estimation as words, the set of all words form the vocabulary and each Raga as a class. We detail our approach to raga recognition as below.

\subsection{Preprocessing}
\subsubsection{Deep Learning based audio source separation}

Audio recordings available as part of CMD accurately represent live ICM concerts with vocalists accompanied by various string, wind, and percussion instruments. However, for the purposes of raga recognition, analyzing the vocals is sufficient. We believe that the presence of other elements in the audio interferes with the performance of the classifier and hence we include audio source separation as a preprocessing step. We use Mad-TwinNet \cite{drossos2018mad}, a deep neural network based approach capable of recovering vocals from a single channel track comprised of vocals and instruments. It uses a Masker Denoiser architecture combined with twin networks (a technique used to regularize recurrent generative networks). We use the pre-trained model made publicly available by the authors.

\subsubsection{Pitch tracking}

Our model analyzes the melody content of audio excerpts to perform Raga recognition, hence pitch tracking is an essential preprocessing step. To perform pitch tracking, we use the python API \cite{jadoul2018introducing} for Praat  \cite{boersma2002praat} which is an open source software for the analysis of speech and sound.

\subsubsection{Tonic Normalization and note quantization}
ICM utilizes a relative scale that is based on the Tonic note of the performer. Since the Tonic can vary in different renditions of the same raga, recognizing and accounting for the tonic note is extremely important. 
\cite{gulati2016phrase} have made numerous features, including Tonic of audio excerpts, available as part of CMD. We normalize the pitch tracked array of every audio in CMD by using the following:
\begin{equation}\label{tonic norm 1}
f_n = 1200*log_{2}(f/T)
\end{equation}
where f is the frequency after pitch tracking (in Hz), T is the frequency of the Tonic (in Hz) and $f_{n}$ is the tonic normalized frequency. The above expression results in 100 cents in a half step, for instance, E and F. Empirically we observe that it is sufficient to have just 5 levels in a half step. However, we leave this as a hyperparameter on the model. Hence the expression for obtaining a tonic normalized note is given by:
\begin{equation}\label{tonic norm 2}
f_p = \mathrm{round}(1200*log_{2}(f/T)*(k/100))
\end{equation}
where k is number of desired levels in a half step. For instance, $k=5$ when number of desired levels between two consecutive notes are 5.

\subsubsection{Random sampling}

The pitch tracked sequences of audio excerpts from CMD have a length of at least $5*10^{5}$ steps. It is impractical to train a Recurrent neural network on sequences with such length as it increases training time considerably. Also, the network would struggle to retain information over such large sequences. Hence, we train the network on \emph{subsequences}, which are smaller sequences randomly sampled from the pitch tracked audio excerpt. It is random in that the start index is randomly selected. We sample $N_{r}$ number of times from every audio excerpt. Length of the subsequence is chosen carefully as it impacts the models training considerably. We observe that smaller sequences tend to confuse the model. An ideal value for the length of the subsequence is 4000-5000 steps.

\subsubsection{Choosing an appropriate value for $N_r$}
Eq (3), determined empirically, gives the appropriate value for $N_r$, based on the length of the subsequence $L_r$ and maximum of the set of lengths of all pitch tracked audio excerpts $L_{max}$.
\begin{equation}\label{eqn3}
N_r = \ceil{2.2 * L_{max} / L_r}
\end{equation}

\subsection{Model Architecture}
At the core of our model (refer figure 1) is the LSTM \cite{hochreiter1997long,45500} based recurrent neural network, which is a popular choice for sequence classification and sequence to sequence learning tasks. We experiment with various configurations for the LSTM block and emperically find that a single LSTM layer with a hidden size of 768 works the best. The word embedding layer, which appears prior to the LSTM layer, converts each note that appears in the subsequence into a 128-dimensional vector. The LSTM block is followed by an attention layer. Our implementation closely follows soft alignment as described in \cite{bahdanau2014neural}. The attention layer is followed by two densely connected layers, the first of which has 384 hidden units and the second one has hidden units equal to the number of classes represented in the training dataset. This is followed by a softmax layer, which converts the output of the final dense layer into a probability distribution over the given number of classes i.e, a vector of length equal to the number of classes and sums to one.

\subsection{Model training and Optimization}

We train the model for over 50 epochs, with a batch size of 40 (i.e, 40 subsequences are fed to model at each training step) and an initial learning rate of 0.0001. We employ dropout \cite{srivastava2014dropout} in the dense layer and batch normalization \cite{ioffe2015batch} after the LSTM layer to reduce over fitting.

\subsubsection{Loss Function}
Since we are dealing with a classification problem with number of classes greater than 2, the categorical cross entropy is the best suited loss function. The categorical cross entropy loss (CCE) is computed as follows:
\begin{equation}\label{eqn3}
\mathrm{CCE} = - \sum_{i=1}^{N} y_{i,j} * log(p_{i})
\end{equation}
Where N is the number of classes, $y_{i,j}$ is an indicator function which is 1 only when $i=j$, j is the training label for the given subsequence, $p_i$ is the $i^{th}$ entry of the probability vector (which is the output of the model).
\subsubsection{Optimization}
Random sampling dramatically increases the size of the data set (the size would be 480*$N_r$, where 480 is the number of recordings in CMD). This makes it hard to fit all these samples on a single computer. Hence we use the Distributed asynchronous stochastic gradient descent training algorithm \cite{dean2012large}. We use the Adam \cite{kingma2014adam} optimizer to update the gradients during training.

%% file: Section3.tex
\begin{figure}
 \centerline{
 \includegraphics[width=\columnwidth,keepaspectratio]{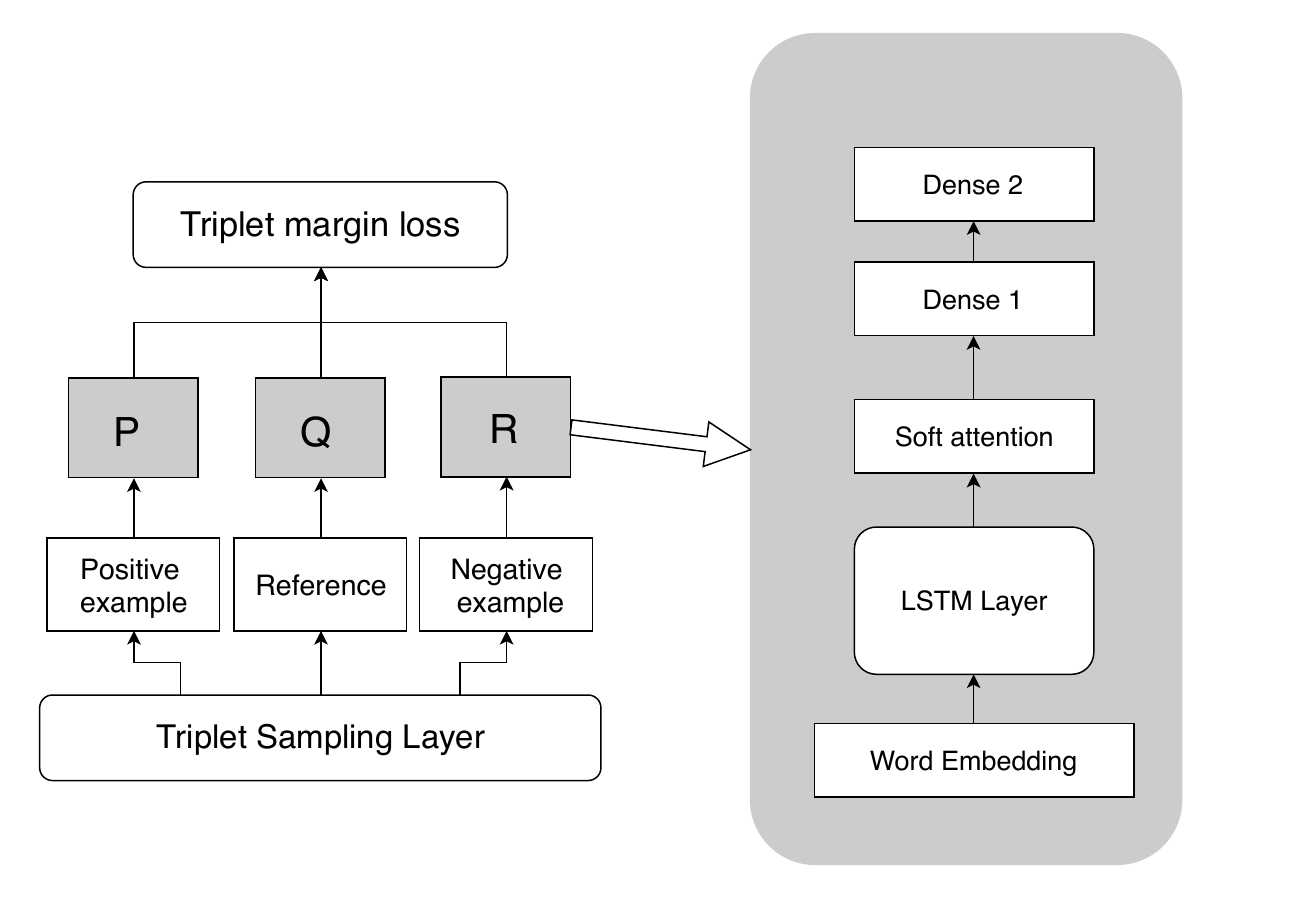}}
 \caption{Schematic diagram for the sequence ranking algorithm. P, Q and R are the copies of the same model and hence have the same architecture.}
 \label{Figure2}
\end{figure}

\section{Sequence Ranking and Retrieval (SRGM2)}

A fully trained sequence ranking model will be able to create feature embeddings such that the L2 distance between the feature embeddings of a pair of sequences is directly proportional to how similar the sequences are. In this work we fine tune a pre-trained Raga recognition model (SRGM1, from the previous section) using the triplet margin loss and evaluate its ability in retrieving subsequences similar to query subsequence. The demonstrated ability to retrieve similar sounding musical pieces is a unique and highly useful feature of our method.

\subsection{Preprocessing}

The preprocessing steps used in the training of SRGM1 apply to sequence ranking. We start with audio source separation followed by pitch tracking, tonic normalization and finally random sampling.

\subsection{Model architecture}

The network architecture of the sequence ranking model is shown in \ref{Figure2}. P, Q, and R are modified copies of pre-trained SRGM1 network (Note: the diagram shows P, Q, and R as different blocks. However, in practice, the triplets are passed through the same network one after the other). The modifications being, 1) absence of softmax activation at the end of the network 2) Dense 2 is altered to have 600 hidden units in place of 40 (or 10 depending on the dataset it was trained on).

\subsection{Triplet sampling}

The triplet sampling layer supplies the model with triplets such that one pair is similar to each other and the other pair is dissimilar. We consider two sequences to be similar if they are from the same raga and are dissimilar if they are from different ragas. The procedure for sampling triplets is as follows:

\begin{itemize}
    \item From the collection of all subsequences, sample a subsequence $P_{ref}$ randomly (uniform random). 
    \item Let $P_{ref}$ belong to raga $R_i$
    \item Select another phrase $P_{+}$ from the same raga $R_i$ randomly (uniform random) such that $P_{+} != P_{ref}$
    \item Randomly (uniform random) choose another Raga $R_j$ such that $R_j != R_i$ from the set of all Ragas $R$
    \item Sample $P_{-}$ from Raga $R_j$ randomly (uniform random).
\end{itemize}

\subsection{Model training and Optimization}

Similar to the training of SRGM1, we employ dropout on the dense layers of the model. Since we are fine-tuning SRGM1, we do not have to train the word embeddings and the LSTM layers. Triplet sampling can be offline (before training) or online (during training). Although it reduces training time, offline sampling is extremely inefficient as it requires a lot of memory (RAM). Hence we sample triplets as and when required. The sequence ranking model is optimized based on the triplet margin loss, given by: 
\begin{equation}\label{eqn3}
\mathcal{L} = min(D(\mathcal{P}_+, \mathcal{P}_{ref}) - D(\mathcal{P}_-, \mathcal{P}_{ref}) + M, 0)
\end{equation}
where $\mathcal{P}$ is the feature embedding obtained for any input P using the model. D(.) is a distance function (the most common choice is the Euclidean distance) and M is the margin (common choice 1). Similar to SRGM1, we use distributed asynchronous stochastic gradient descent for training the model with Adam optimizer.

%% file: Section4.tex
\section{Dataset, Inference and Evaluation methodology}

\subsection{Dataset}

In this work, we use the Carnatic Music Dataset \cite{gulati2016phrase} made available by the Comp Music group. The Comp Music data set consists of high-quality recordings of live ICM concerts. This dataset comprises of 124 hours of 480 commercially available recordings that are stored in the mp3 format. 

The dataset contains 12 full-length recordings per Raga and features 40 different Ragas (hence 480 recordings). These recordings feature numerous artists, a wide variety of compositions and a diverse set of Ragas. The ragas are diverse in terms of their melodic attributes (refer to Section 1). Therefore, we believe that training and evaluating the model on this dataset would allow us to assess how well the model would perform in near real-world scenarios.

\begin{figure}
 \centerline{
 \includegraphics[width=\columnwidth]{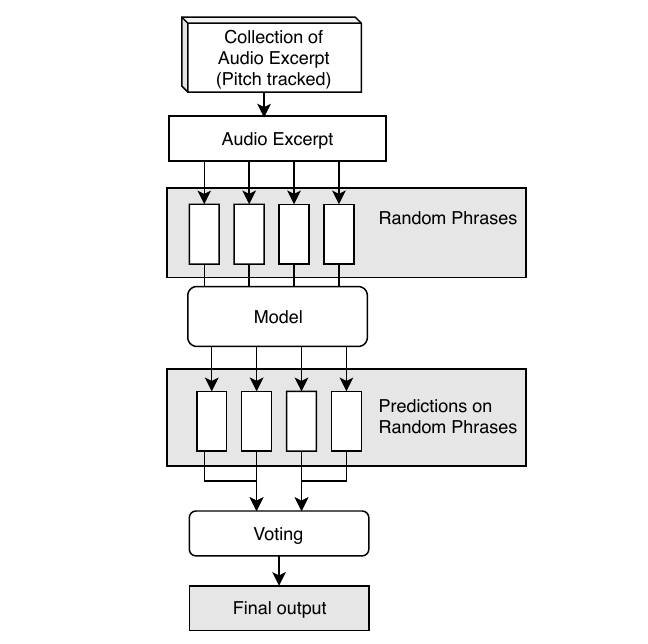}}
 \caption{Figure gives an overview of the inference process for SRGM1 as described in Section 5.3.1}
 \label{Figure3}
\end{figure}

\subsection{Comparison with prior works}

We compare our results with three prior works, \cite{gulati2016phrase, gulati2016time, chordia2013joint} of which \cite{gulati2016time} is the current state of the art for the Raga recognition task on the CMD dataset and \cite{gulati2016phrase} is the current state of the art for Raga recognition on the 10 Raga subset of CMD dataset. Gulati et al. \cite{gulati2016time} provides a summary of the performance of all of these the three methods on the CMD dataset. To ensure a fair comparison, we follow the evaluation strategy as provided in \cite{gulati2016time} (described in 3.3.1).

\subsection{Inference}
\subsubsection{SRGM1}
Although the model is trained on subsequences, during inference, we are interested in determining the Raga of the audio as a whole. Thus, follow the procedure as shown in figure \ref{Figure3}. First, we split the audio excerpt into numerous subsequences and obtain the predictions for each of these. We then perform voting to determine the majority class. Note, that if the majority class has less than 60\% of the votes we label the classification as incorrect.

\subsubsection{SRGM2}

For SRGM2, we are interested mostly in the sequences the model retrieves rather than inference on the audio as a whole. However, if one desires to make inference on the audio as a whole, we can adopt the procedure described in section 3.3.1. 

\subsection{Evaluation strategy}

\subsubsection{Evaluation strategy for SRGM1}
CMD consists of 12 recordings per Raga. Since we train the model on subsequences, we have a total of 12*$N_r$ subsequences in every class. Therefore, this is a balanced classification problem and accuracy is a suitable metric for evaluation. Authors in \cite{gulati2016time} use leave-one-out cross-validation strategy for evaluation where one of the 12 recordings for every Raga is used as test data and the remaining 11 are used as the training data. We adopt the same approach to evaluate our model as it enables us to make fair comparison of our results with previous works. We also present a confusion matrix for the observed results to further investigate the performance of SRGM1. 

Gulati et al. \cite{gulati2016phrase} use stratified 12 fold cross-validation. To be able to compare our results to theirs on the 10 Raga subset of CMD, we adopt the same strategy. As an alternative evaluation strategy, we hold out 5 of the 12 recordings from every class as the test set and train the model on the rest.

\subsubsection{Evaluation strategy for SRGM2}

A suitable evaluation metric for SRGM2, a sequence ranking model, is the ``precision at top-k'' measure which is popular score used to evaluate information retrieval systems. Precision at top k is defined as the proportion of retrieved items in the top-k set that are relevant to the query. We consider a retrieved subsequence to be relevant when it belongs to the same Raga as the query subsequence. The expression for precision at top-k hence becomes :
\begin{equation}\label{eqn3}
P_{k} = \frac{\sum_{i=1}^{k} I_{y_i, c}}{k}
\end{equation}
\begin{equation}\label{eqn3}
P_{avg} = \frac{\sum_{i=1}^{n} P_{k_i}}{n}
\end{equation}
Where $P_k$ is precision at top k evaluated on one query sample, $I_{y_i}$ is an indicator function which is 1 only when $y_i$ which is the Raga of the $i^{th}$ retrieved sample is same as $c$ which is the Raga of the query sample. $P_{avg}$ is the top-k precision averaged over the complete dataset.

%% file: Section5.tex
\section{Results and discussion}

\begin{table}
 \begin{center}
 \begin{tabular}{|l|l|l|}
  \hline
  \textbf{Method} & \textbf{CMD-10 Ragas} & \textbf{CMD-40 Ragas} \\
  \hline
  SRGM1  & 95.6\% & 84.6\% \\
  \hline
  SRGM1 Ensemble  & \textbf{97.1}\% & \textbf{88.1}\% \\
  \hline
  $\mathcal{M}_{F}$ & - & 81.5 \% \\
  \hline
  $\mathcal{M}_{KL}$ & - & 86.7 \% \\
  \hline
  $\mathcal{M}_{B}$ & - & 86.7 \% \\
  \hline
  $\mathcal{E}_{VSM}$ & 91.7 \% & 68.1 \% \\
  \hline
  $\mathcal{E}_{PCD}$ & 82.2 \% & 73.1 \% \\
  \hline
 \end{tabular}
\end{center}
 \caption{Table summarizes performances of various models on the CMD dataset and its 10 Raga subset. Refer to Section 6.1}
 \label{tab:table1}
\end{table}

We present a summary of results in table \ref{tab:table1} where we compare the performance of our model with \cite{gulati2016time}, \cite{gulati2016phrase} and \cite{chordia2013joint}. $\mathcal{M}_F$, $\mathcal{M}_{KL}$ and $\mathcal{M}_B$ represent 3 variations of TDMS \cite{gulati2016time} which uses euclidean distance, KL divergence and Bhattacharya distance respectively. $\mathcal{E}_{VSM}$ represents vector space model \cite{gulati2016phrase} for Raga recognition while $\mathcal{E}_{PCD}$ represents pitch class distribution based method presented in \cite{chordia2013joint}. 

We observe that our method performs better than the current state-of-the-art on both 10 Raga subset of CMD and the CMD as a whole. We obtain an improvement of ~ 6 \% on the 10 Raga subset of CMD and improvement of ~ 2\% on CMD using the SRGM1-Ensemble model. The Ensemble model was created using 4 sets of weights for the same model architecture. For each test sample, we obtain output from each of the 4 models and then combine them by summing the log of the outputs of the model to obtain the final prediction vector for that sample.

\begin{figure}
 \centerline{
 \includegraphics[width=\columnwidth]{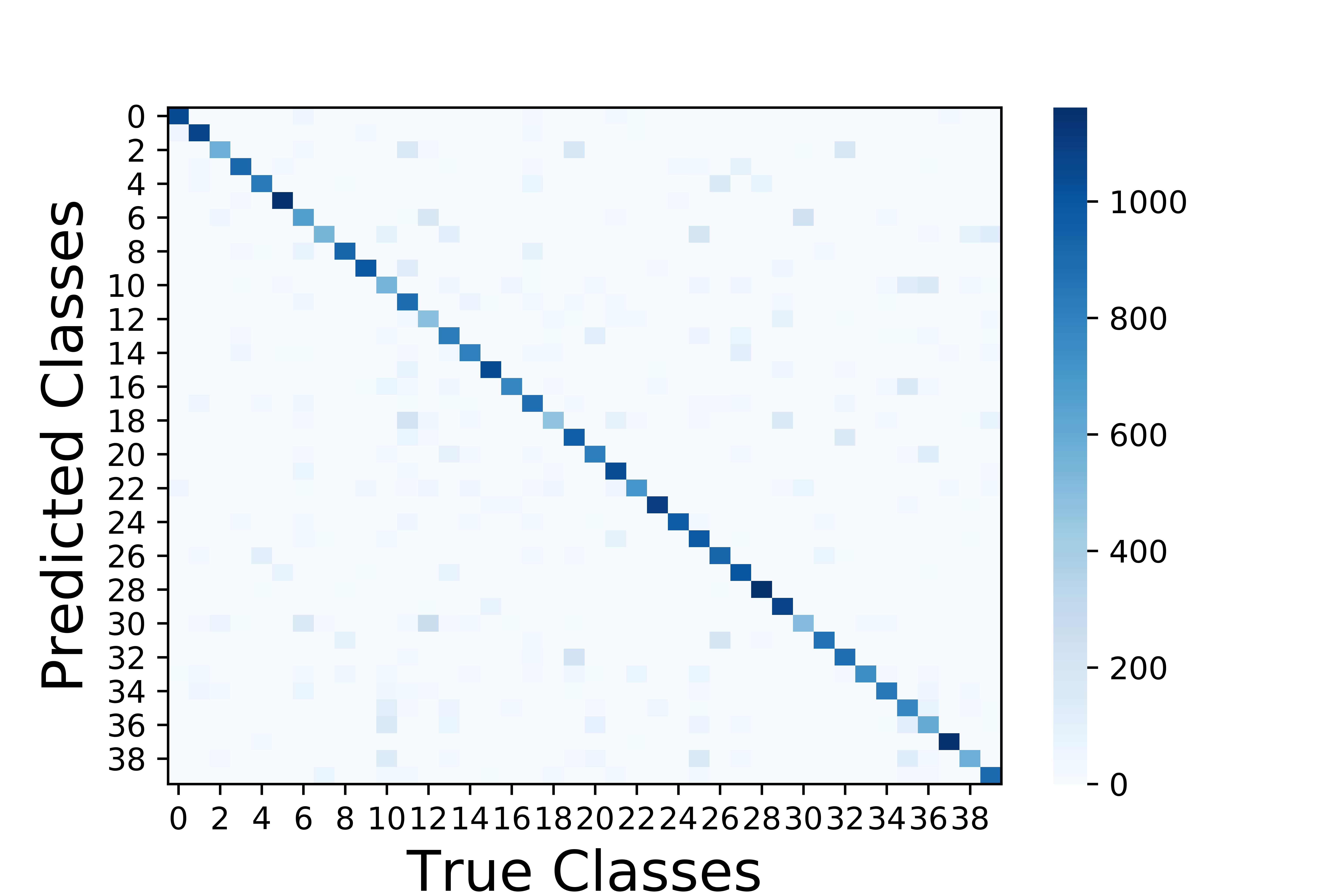}}
 \caption{Figure shows the confusion matrix for the predictions obtained using SRGM1.}
 \label{Figure5}
\end{figure}

The confusion matrix for SRGM1 trained on CMD (40 Ragas) is as shown in figure \ref{Figure5}. Note that the confusion matrix is presented for the results obtained on a subsequence level and not on the audio excerpt level i.e, these are results before performing inference (refer to Section 4.3.1). There are no patterns apparent from the confusion matrix. However, on closely observing the samples on which the model performs poorly, we see that these samples have very little information, which can be attributed to events like to long pauses during the performance, sustaining a particular note for a prolonged interval, etc.

\begin{table}[]
\begin{tabular}{|c|c|c|c|}
\hline
\textbf{\begin{tabular}[c]{@{}c@{}}Subsequence\\ length\end{tabular}} & \textbf{\begin{tabular}[c]{@{}c@{}}Num epochs\\ to converge\end{tabular}} & \textbf{\begin{tabular}[c]{@{}c@{}}Time per\\ epoch in sec\\ (Wall Time)\end{tabular}} & \textbf{\begin{tabular}[c]{@{}c@{}}Hold out\\ test set\\ accuracy\end{tabular}} \\ \hline
500  & 12 & 32  &  88.86\\ \hline
1500 & 6  & 58.4  &  95.5\\ \hline
3000 & 5  & 120.1  &  95.63\\ \hline
6000 & 3  & 241.5  &  97.34 \\ \hline
\end{tabular}\label{tab:table2}
\caption{Summary of our findings for the study on variation of model training and performance based on subsequence length}
\end{table}

\begin{figure}
\centering
 \centerline{
 \includegraphics[width=10.5cm,keepaspectratio]{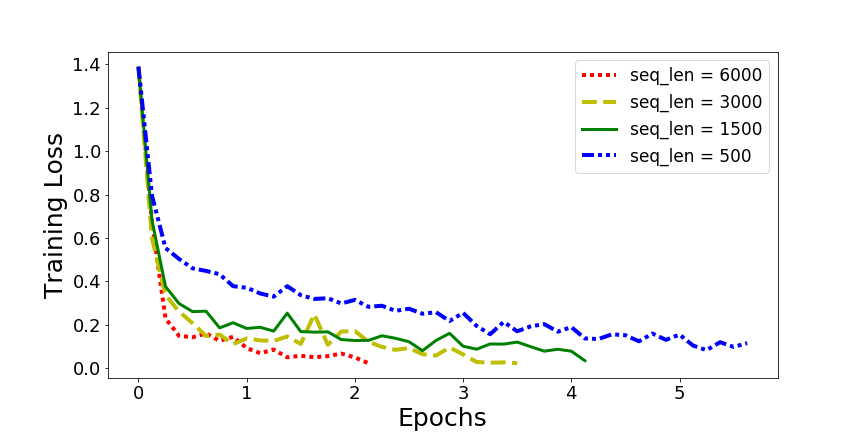}}
 \caption{The above graph depicts variation in the "training loss vs epochs" plot with changing subsequence length.}
 \label{Figure6}
\end{figure}

To further investigate the effect of subsequence length on the training of a model, we devise an experiment by using a 4 Raga subset of CMD. We use 36 recordings as the training set and 12 recordings as the test set. We train the network until the categorical cross entropy loss reduces to 0.02. Figure \ref{Figure6} shows a plot of the training loss vs the number of epochs. It is clearly visible that a model that is trained on subsequences of length 6000 converges in lesser number of epochs and has a smooth descent while the model using subsequences of length 500 makes the training very noisy and the network takes as long as 12 epochs to attain the same value of the loss. A summary of our findings has been tabulated in table 2.

\begin{table}[]
\centering
\begin{tabular}{|l|c|c|}
\hline
\textbf{Metric} & \begin{tabular}[c]{@{}c@{}}top-30 \\ precision\end{tabular} & \begin{tabular}[c]{@{}c@{}}top-10 \\ precision\end{tabular} \\ \hline
\textbf{Score} & 81.83 \% & 81.68 \% \\ \hline
\end{tabular}
\caption{Summary of performance of sequence ranking on the top-10 and top-30 precision metrics}
\end{table}

We evaluate the sequence ranking model on the metrics described in section 5.4.2, namely top-10 precision and top-30 precision. Our findings have been summarized in table 3. SRGM2 obtains a top-30 precision of 81.83 \% and top-10 precision of 81.68 \%. We also conduct a qualitative analysis of the retrieved samples by inspecting various aspects like the range of svaras (notes) observed in the query to that in the retrieved sample, checking for similar note progressions, etc. We observe that the sequences retrieved by the model are similar to the ones in the query. We attach several samples of query subsequences and retrieved sequences as part of the supplementary material.

\section{Conclusion and future work}

In this work, we proposed a novel method to address the problem of Raga recognition. Through various experiments and validation, we are able to demonstrate the capability of our approach in tackling the problem of raga recognition. We also introduce sequence ranking as a new sub-task of raga recognition. We present numerous query - retrieved subsequence pairs (as part of supplementary material) which demonstrates the effectiveness of this approach to mine databases for similar sequences.

We believe that deep learning based approaches have tremendous scope in MIR pertinent to ICM. As part of future work, we would like to further explore the idea of sequence ranking as we feel it can be used for tasks like joint recognition of tonic and Raga. An exciting future research direction would be to explore the possibility of using deep learning for generative modeling tasks in ICM. It would definitely be interesting to see if deep learning models can replicate the intricate improvisation aspects of ICM.

%% file: main.bbl
\begin{thebibliography}{10}

\bibitem{bahdanau2014neural}
Dzmitry Bahdanau, Kyunghyun Cho, and Yoshua Bengio.
\newblock Neural machine translation by jointly learning to align and
  translate.
\newblock {\em arXiv preprint arXiv:1409.0473}, 2014.

\bibitem{balkwill1999cross}
Laura-Lee Balkwill and William~Forde Thompson.
\newblock A cross-cultural investigation of the perception of emotion in music:
  Psychophysical and cultural cues.
\newblock {\em Music perception: an interdisciplinary journal}, 17(1):43--64,
  1999.

\bibitem{boersma2002praat}
Paul Boersma et~al.
\newblock Praat, a system for doing phonetics by computer.
\newblock {\em Glot international}, 5, 2002.

\bibitem{chordia2013joint}
Parag Chordia and Sertan {\c{S}}ent{\"u}rk.
\newblock Joint recognition of raag and tonic in north indian music.
\newblock {\em Computer Music Journal}, 37(3):82--98, 2013.

\bibitem{dean2012large}
Jeffrey Dean, Greg Corrado, Rajat Monga, Kai Chen, Matthieu Devin, Mark Mao,
  Andrew Senior, Paul Tucker, Ke~Yang, Quoc~V Le, et~al.
\newblock Large scale distributed deep networks.
\newblock In {\em Advances in neural information processing systems}, pages
  1223--1231, 2012.

\bibitem{dighe2013scale}
Pranay Dighe, Parul Agrawal, Harish Karnick, Siddartha Thota, and Bhiksha Raj.
\newblock Scale independent raga identification using chromagram patterns and
  swara based features.
\newblock In {\em International Conference on Multimedia and Expo Workshops
  (ICMEW)}, pages 1--4. IEEE, 2013.

\bibitem{dighe2013swara}
Pranay Dighe, Harish Karnick, and Bhiksha Raj.
\newblock Swara histogram based structural analysis and identification of
  indian classical ragas.
\newblock In {\em International Society for Music Information Retrieval
  Conference}, pages 35--40, 2013.

\bibitem{drossos2018mad}
Konstantinos Drossos, Stylianos~Ioannis Mimilakis, Dmitriy Serdyuk, Gerald
  Schuller, Tuomas Virtanen, and Yoshua Bengio.
\newblock Mad twinnet: Masker-denoiser architecture with twin networks for
  monaural sound source separation.
\newblock In {\em International Joint Conference on Neural Networks (IJCNN)},
  pages 1--8. IEEE, 2018.

\bibitem{dutta2015raga}
Shrey Dutta and Hema~A Murthy.
\newblock Raga verification in carnatic music using longest common segment set.
\newblock In {\em International Society for Music Information Retrieval
  Conference}, volume~1, pages 605--611, 2015.

\bibitem{ganti2013bollywood}
Tejaswini Ganti.
\newblock {\em Bollywood: A guidebook to popular Hindi cinema}.
\newblock Routledge, 2013.

\bibitem{gulati2016phrase}
Sankalp Gulati, Joan Serra, Vignesh Ishwar, Sertan Sent{\"u}rk, and Xavier
  Serra.
\newblock Phrase-based r{\=a}ga recognition using vector space modeling.
\newblock In {\em IEEE International Conference on Acoustics, Speech and Signal
  Processing (ICASSP)}, pages 66--70. IEEE, 2016.

\bibitem{gulati2016time}
Sankalp Gulati, Joan Serr{\`a}~Juli{\`a}, Kaustuv~Kanti Ganguli, Sertan
  Sent{\"u}rk, and Xavier Serra.
\newblock Time-delayed melody surfaces for r{\=a}ga recognition.
\newblock In {\em International Society for Music Information Retrieval
  Conference}, 2016.

\bibitem{hochreiter1997long}
Sepp Hochreiter and J{\"u}rgen Schmidhuber.
\newblock Long short-term memory.
\newblock {\em Neural computation}, 9(8):1735--1780, 1997.

\bibitem{ioffe2015batch}
Sergey Ioffe and Christian Szegedy.
\newblock Batch normalization: Accelerating deep network training by reducing
  internal covariate shift.
\newblock {\em arXiv preprint arXiv:1502.03167}, 2015.

\bibitem{jadoul2018introducing}
Yannick Jadoul, Bill Thompson, and Bart De~Boer.
\newblock Introducing parselmouth: A python interface to praat.
\newblock {\em Journal of Phonetics}, 71:1--15, 2018.

\bibitem{kingma2014adam}
Diederik~P Kingma and Jimmy Ba.
\newblock Adam: A method for stochastic optimization.
\newblock {\em arXiv preprint arXiv:1412.6980}, 2014.

\bibitem{koduri2012raga}
Gopala~Krishna Koduri, Sankalp Gulati, Preeti Rao, and Xavier Serra.
\newblock R{\=a}ga recognition based on pitch distribution methods.
\newblock {\em Journal of New Music Research}, 41(4):337--350, 2012.

\bibitem{koduri2014intonation}
Gopala~Krishna Koduri, Vignesh Ishwar, Joan Serr{\`a}, and Xavier Serra.
\newblock Intonation analysis of r{\=a}gas in carnatic music.
\newblock {\em Journal of New Music Research}, 43(1):72--93, 2014.

\bibitem{krishna2011identification}
A~Srinath Krishna, PV~Rajkumar, KP~Saishankar, and Mala John.
\newblock Identification of carnatic raagas using hidden markov models.
\newblock In {\em International Symposium on Applied Machine Intelligence and
  Informatics (SAMI)}, pages 107--110. IEEE, 2011.

\bibitem{krishna2012carnatic}
TM~Krishna and Vignesh Ishwar.
\newblock Carnatic music: Svara, gamaka, motif and raga identity.
\newblock In {\em Proceedings of the 2nd CompMusic Workshop}. Universitat
  Pompeu Fabra, 2012.

\bibitem{madhusdhan2018tonic}
Sathwik~Tejaswi Madhusdhan and Girish Chowdhary.
\newblock Tonic independent raag classification in indian classical music.
\newblock 2018.

\bibitem{45500}
Christopher Olah.
\newblock Understanding lstm networks, 2015.

\bibitem{shetty2009raga}
Surendra Shetty and KK~Achary.
\newblock Raga mining of indian music by extracting arohana-avarohana pattern.
\newblock {\em International Journal of Recent Trends in Engineering},
  1(1):362, 2009.

\bibitem{sridhar2009raga}
Rajeswari Sridhar and TV~Geetha.
\newblock Raga identification of carnatic music for music information
  retrieval.
\newblock {\em International Journal of recent trends in Engineering},
  1(1):571, 2009.

\bibitem{srivastava2014dropout}
Nitish Srivastava, Geoffrey Hinton, Alex Krizhevsky, Ilya Sutskever, and Ruslan
  Salakhutdinov.
\newblock Dropout: a simple way to prevent neural networks from overfitting.
\newblock {\em The Journal of Machine Learning Research}, 15(1):1929--1958,
  2014.

\bibitem{van2013deep}
Aaron Van~den Oord, Sander Dieleman, and Benjamin Schrauwen.
\newblock Deep content-based music recommendation.
\newblock In {\em Advances in neural information processing systems}, pages
  2643--2651, 2013.

\bibitem{wang2014improving}
Xinxi Wang and Ye~Wang.
\newblock Improving content-based and hybrid music recommendation using deep
  learning.
\newblock In {\em Proceedings of the 22nd ACM international conference on
  Multimedia}, pages 627--636. ACM, 2014.

\end{thebibliography}
